# Protein corona critically affects the bio-behaviors of SARS-CoV-2

Yue-wen Yin[1], Yan-jing Sheng[1], Min Wang[1], Song-di Ni,[1] Hong-ming Ding[1,*], Yu-qiang Ma[2,*]

[1] Center for Soft Condensed Matter Physics and Interdisciplinary Research, School of Physical Science and Technology, Soochow University, Suzhou 215006, China.

[2] National Laboratory of Solid State Microstructures and Department of Physics, Collaborative Innovation Center of Advanced Microstructures, Nanjing University, Nanjing 210093, China.

*Email: dinghm@suda.edu.cn, myqiang@nju.edu.cn

## ABSTRACT

The outbreak of the coronavirus disease 2019 (COVID-19) caused by severe acute respiratory syndrome coronavirus-2 (SARS-CoV-2) has become a worldwide public health crisis. When the SARS-CoV-2 enters the biological fluids in the human body, different types of biomolecules (in particular proteins) may adsorb on its surface and alter its infection ability. Although great efforts have recently been devoted to the interaction of the specific antibodies with the SARS-CoV-2, it still remains largely unknown how the other serum proteins affect the infection of the SARS-CoV-2. In this work, we systematically investigate the interaction of serum proteins with the SARS-CoV-2 RBD by the molecular docking and the all-atom molecular dynamics simulations. It is found that the non-specific immunoglobulin (Ig) indeed cannot effectively bind to the SARS-CoV-2 RBD while the human serum albumin (HSA) may have some potential of blocking its infection (to ACE2). More importantly, we find that the RBD can cause the significant structural change of the Apolipoprotein E (ApoE), by which SARS-CoV-2 may hijack the metabolic pathway of the ApoE to facilitate its cell entry. The present study enhances the understanding of the role of protein corona in the bio-behaviors of SARS-CoV-2, which may aid the more precise and personalized treatment for COVID-19 infection in the clinic.

**KEYWORDS:** protein corona, SARS-CoV-2, neutralization, enhanced infection, molecular modeling

## 1. Introduction

The outbreak of COVID-19 disease that is caused by the SARS-CoV-2 has led to more than 102.0 million infections and over 2.2 million deaths in the whole world by Jan 31, 2021 (https://www.who.int/emergencies/diseases/novel-coronavirus-2019/situation-reports/). The spike protein on the viral membrane is the key antigen of SARS-CoV-2, where the receptor binding domains (RBDs) located on the S1 domain of the spike protein can facilitate viral entry via interaction with angiotensin-converting enzyme 2 (ACE2)[1-3].

Due to its crucial role in viral infection, SARS-CoV-2 RBD has become one of the most likely targets for the development of drugs/inhibitors, neutralizing antibodies, and vaccines[4-15]. For example, Panda et al.[4] found that an antiviral inhibitor showed enhanced binding affinity to the RBD, where the postfusion conformation of the trimeric S protein RBD with ACE2 revealed conformational changes associated with PC786 drug binding. Moreover, the antibodies that targeting the RBD of the spike glycoprotein are believed to be a more promising method of treating COVID-2019. Yan et al.[5] isolated two specific human monoclonal antibodies (i.e., CA1 and CB6) from a convalescent COVID-19 patient, and demonstrated potent SARS-CoV-2-specific neutralization activity in vitro against SARS-CoV-2. More recently, Barnes et al.[6] made a more comprehensive study on the neutralization antibodies, where they solved eight new structures of distinct COVID-19 human neutralizing antibodies in complex with the SARS-CoV-2 spike trimer or RBD and more importantly classified these antibodies into categories based on the binding modes.

Actually, in the human body, there are many types of antibodies (i.e., immunoglobulins, Igs) that identify and neutralize foreign objects such as pathogenic bacteria and viruses although these intrinsic antibodies are not specific for the SARS-CoV-2. Apart from the Igs, there are many types of serum proteins in the blood. For example, the human serum albumin (HSA) is the most abundant protein in human blood plasma and constitutes about half of the blood serum protein[16]. Apolipoproteins (mainly consisting of ApoA, ApoB, and ApoE) can bind lipids and transport them in blood, cerebrospinal fluid, and lymph[17]. When the foreigner objects/particles enter into the human body, these serums proteins could quickly adsorb on the foreign particle surface, and form the protein corona, which have been proved to significantly affect the bio-behaviors of the particles[18-25].

Since the size of the virus is in the nanometers, the serum protein may also adsorb on the viral surface and in turn affect its bio-behaviors[26-28]. For example, Ezzat et al.[28] demonstrated respiratory syncytial virus (RSV) and herpes simplex virus type 1 (HSV-1) accumulated by a rich protein corona in the biological fluids, and this protein corona could affect the viral infectivity and immune cell activation.

In this work, with the aim of clarifying the role of protein corona in SARS-CoV-2 infection, we systematically investigate the interaction of three typical serum proteins (i.e., IgM, HSA, and ApoE) with the SARS-CoV-2 RBD by combining the molecular docking, the all-atom molecular dynamics simulations, and the binding energy calculation. Our results show that the HSA proteins could adsorb onto the RBD and hide the antigen site of the RBD while the non-specific IgM cannot. More importantly, the binding to RBD can cause the obviously structural change of the ApoE, which may activate the metabolic pathway of the ApoE and facilitate the SARS-CoV-2 entry into specific cells.

## 2. Materials and methods

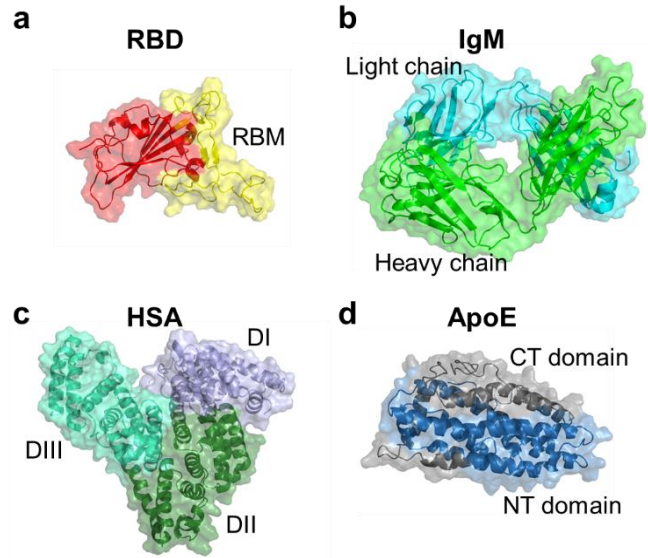

**Figure 1.** The crystal structure of the SARS-CoV-2 RBD and three typical serum proteins used in this work. (a) SARS-CoV-2 RBD, (b) Immunoglobulin M, (c) Human serum albumin, and (d) Apolipoprotein E.

*2.1 The structure of the serum proteins and SARS-CoV-2 RBD*

As shown in Figure 1, the SARS-CoV-2 RBD (PDB ID: 6M0J)[29] and three different serum proteins, namely IgM (PDB ID: 1DEE)[30], HSA (PDB ID: 4K71)[31], and ApoE (PDB ID: 2L7B)[32], were considered here. The RBD is the main functional region of SARS-CoV-2 spike protein interacting with ACE2, especially the receptor binding motif (RBM), which contains almost all the key residues related to the binding between SARS-CoV-2 and ACE2. The IgM is a heterodimeric protein composed of two heavy chains and two light chains, which comprises two Fab fragments and one single Fc fragment[33]. Since the two Fab fragments are the same and the Fc fragment usually does not bind to the antigen, here for the sake of simplicity, we just considered one Fab segment of IgM. The HSA is a helical protein with turns and extended loops, and resembles a heart shape. It consists of three domains, I (residues 1–195), II (196–383), and III (384–585)[34]. The ApoE is folded into two separate domains with N-terminal (residues 1 to 191) containing an anti-parallel four-helix bundle and C-terminal (residues 192 to 299) forming a separately folded domain that interacts with the four-helix bundle[35].

## 2.2 Molecular docking

The complex structures between the SARS-CoV-2 RBD and the serum proteins were first predicted by the hybrid protein-protein docking approach, HDOCK[36-38]. Since the SARS-CoV-2 RBD and IgM-Fab were part of the spike protein and IgM protein, respectively, the residues that may be shielded by the other part of the proteins were excluded in the docking process. For each complex, the HDOCK algorithm can predict 100 binding conformations that were sorted according to the binding docking scores. However, previous studies[39-41] indicated that the performance of the docking scores on the ranking ability was just moderate and even very poor in some cases; while the binding energy calculated by the molecular mechanics/Poisson-Boltzmann surface area (MM/PBSA) could perform much better. Thus, after the molecular docking, we then calculated the binding energy for the top 30 conformations (predicted by the HDOCK) with the MM/PBSA method. Finally, we selected three typical structures for each complex based on the binding energy and the cluster analysis as the initial conformations for the following MD simulation.

## 2.3 All-atom molecular dynamics (MD) simulation.

The all-atom molecular dynamics (MD) simulations were performed based on the above selected structures. All the MD simulations were carried by using GROMACS 2019.03 package[42] with Amber ff14sb force field[43]. Each complex was solvated with TIP3P water[44], with the LINCS algorithm to constraint all bonds involving hydrogen atoms[45]. The NaCl was used to neutralize system, and the concentration was set at 0.15M. The distances between the surfaces of the box and the complex atoms were set to $\geq 15$Å. The initial system was firstly energy-minimized twice by the steepest descent method before and after filling the water until the convergence was reached. Then each system was subjected to a 500 ps pre-equilibration process in the NVT and NPT ensemble, separately, with all the heavy atoms of proteins harmonically constrained by 1000 kJ*mol$^{-1}$*nm$^{-2}$. The temperature was controlled at 310 K by the V-rescale thermostat with a time constant of 0.2 ps and the pressure was kept at 1 atm by the Parrinello-Rahman barostat with a time constant of 2.0 ps. Finally, 100 ns free NPT simulations were performed for each system. In the simulations, the particle mesh Ewald (PME) algorithm was used when calculating the long-range electrostatic interactions[46], and the Lennard-Jones (LJ) interactions were cut off at a distance of 1.0 nm. The periodic boundary conditions were adopted in all three directions.

*2.4 Binding energy calculation.*

In the binding energy calculation, every interval of 100 ps was sampled in the last 10 ns, namely 100 frames in each system were used to calculate the binding energy via the MM/PBSA method[47-50]. All the MM/PBSA calculations were performed by using the modified shell script gmx_mmpbsa (https://jerkwin.github.io/gmxtool). The typical snapshots for the protein structures were prepared using PyMOL (http://www.pymol.org/).

## 3. Results and discussion

*3.1 Interaction of Immunoglobulin M (IgM) with SARS-CoV-2 RBD*

We first investigated the interaction between the non-specific IgM and SARS-CoV-2 RBD, where three different binding modes were observed (Figure 2 and Figure S1). Model-1 was the typical neutralization mode that the IgM bound to the RBD with its antigen-binding site (i.e., the variable regions in the light chain and the heavy chain). Model-2 was another neutralization mode, but the binding site in RBD was changed. As shown in Figure 2d, the binding energy in the two models was about -20 kcal/mol and -30 kcal/mol, respectively, which was much weaker than the binding energy of the specific antibody to the SARS-CoV-2 RBD (e.g., CB6~-37 kcal/mol, CR3022~-34 kcal/mol)[50]. We also observed an unusual binding mode namely model-3, where the RBD was mainly bound to the junction between the constant region and the variable region of the heavy chain. Due to the hollow structure of the junction, the contact surface area (CSA) between the RBD and the IgM was much larger than that in previous two models (Figure 2f). Nevertheless, there were many polar residues at the binding site (Figure S1e), leading to a high solvation energy, thus the binding energy was also weak (~-22 kcal/mol) in this case.

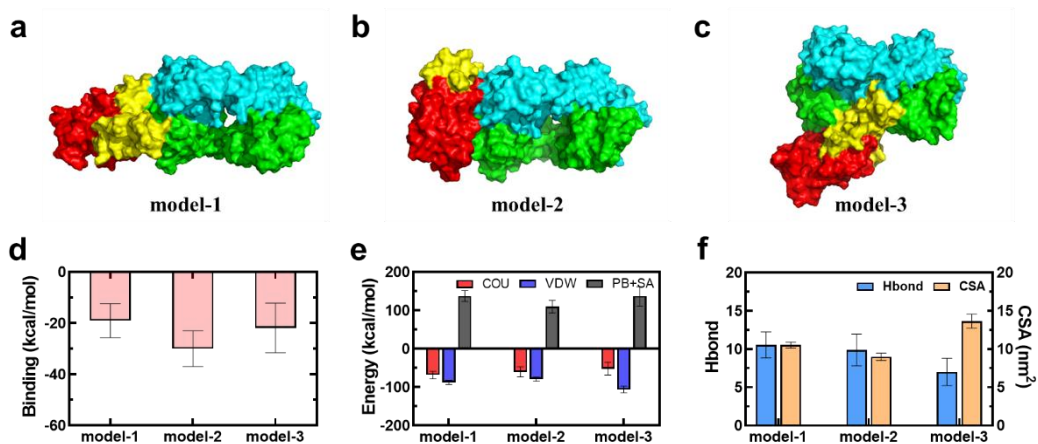

**Figure 2.** Interaction of IgM with SARS-CoV-2 RBD. Final snapshots illustrating the three typical modes in the IgM-RBD interaction (a-c) and corresponding binding energies (d), energy decomposition of the binding energy (e), the number of hydrogen bonds and contact surface areas (f) between the IgM and RBD in the three modes. The contact surface area (CSA) is denoted as: $CSA=(SASA_{IgM} + SASA_{RBD} – SASA_{com})/2$, where SASA stands for the solvent accessible surface area and the footnote 'com' indicate the complex of the IgM and RBD.

Due to the binding of IgM, the epitope in RBD (that binds to ACE2) might be occupied. To quantitatively depict the effect of the IgM binding on the epitope in RBD, we analyzed the buried surface area (BSA)[51] and ΔRMSF[52] of the residues in the epitope (that play important roles in ACE2-RBD interactions). As shown in Figure 3a, the BSA of most residues was positive, namely the solvent accessible surface area (SASA) of the key residues in the epitope became smaller in model-1 and model-3. Moreover, Figure 3b shows that most of ΔRMSF was negative in these two models, meaning that the flexibility of the key residues was reduced and the activity of these residues was weakened. The above results indicated that the epitope was indeed occupied by the IgM in model-1 and model-3. While in model-2, both of the BSA and the ΔRMSF changed very little, indicating that the epitope was not affected by the binding of IgM. Although the IgM did not directly block the ACE2-RBD interaction (similar to the antibody CR3022 reported by Wilson and coworkers)[53], it may also have great impact on the infection of the SARS-CoV-2 (e.g., the clash with the N-terminal domain or the S2 domain may affect the conformation change of the spike protein). Therefore, model-2 may be another useful neutralization model. However, the IgM here was not the specific antibody for SARS-CoV-2 and the binding energy was weak. Thus, the neutralization of the non-specific IgM to RBD was limited although different neutralization modes were observed here.

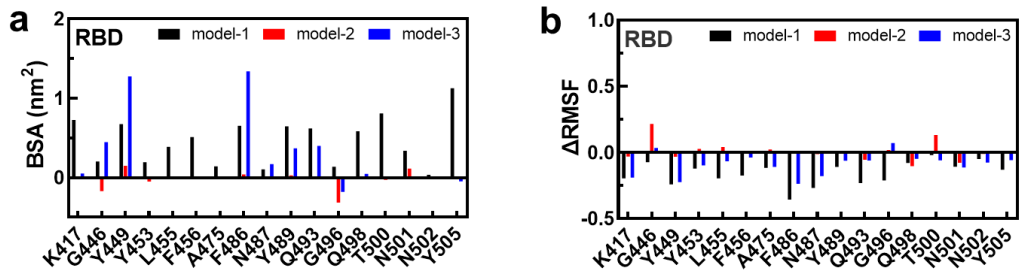

**Figure 3.** The buried surface area (BSA) (a) and the root mean square fluctuation (RMSF) difference parameter ΔRMSF (b) of the epitope site of RBD in the three models. BSA=SASA$_{free}$ – SASA$_{binding}$, ΔRMSF=(RMSF$_{binding}$ – RMSF$_{free}$)/RMSF$_{free}$, where the footnote 'binding' and 'free' indicate the binding state and the free state, respectively. A positive value of BSA indicates that the residue is hidden, while a negative value indicates that the residue becomes exposed after the binding. A positive value of ΔRMSF indicates an increase in the flexibility of residues, while a negative value indicates a decrease in the flexibility.

### 3.2 Interaction of Human serum albumin (HSA) with SARS-CoV-2 RBD

Human serum albumin (HSA) is the most abundant protein in human blood plasma and constitutes over half of the blood serum protein. The adsorption of HSA on the foreign objects can also lead to the immune response[21, 54]. For example, Schöttler et al.[21] found that the pre-adsorption of HSA on the nanoparticle surface can greatly increase the macrophage uptake.

Here, for the purpose of determining the 'neutralization' ability of HSA, we also investigated the binding of the HSA to the SARS-CoV-2 RBD (Figure 4 and Figure S2). As discussed above, HSA contains three homologous helical domains (i.e., DI, DII, and DIII) and each splits into A and B subdomains. Due to the heart-shaped structure of HSA, there was large space for other molecules binding at the junction of the three domains. In this case, we also observed three different modes (see Figure 4a-c), where RBD was bound to the DI-DII junction (model-1), DI-DIII junction (model-2), and DI-DII-DIII junction (model-3), respectively. More importantly, the binding energy was much stronger than that in the case of IgM, and was even stronger than that of RBD-ACE2 (-34 kcal/mol) in the first two models[50]. Figure 4e shows that both of the electrostatic energy and the VDW energy played important roles in the interactions. Besides, as illustrated in Figure S2c-d, there were a lot of hydrophobic contacts in model-1 and model-2, which collectively led to the strong binding energy here.

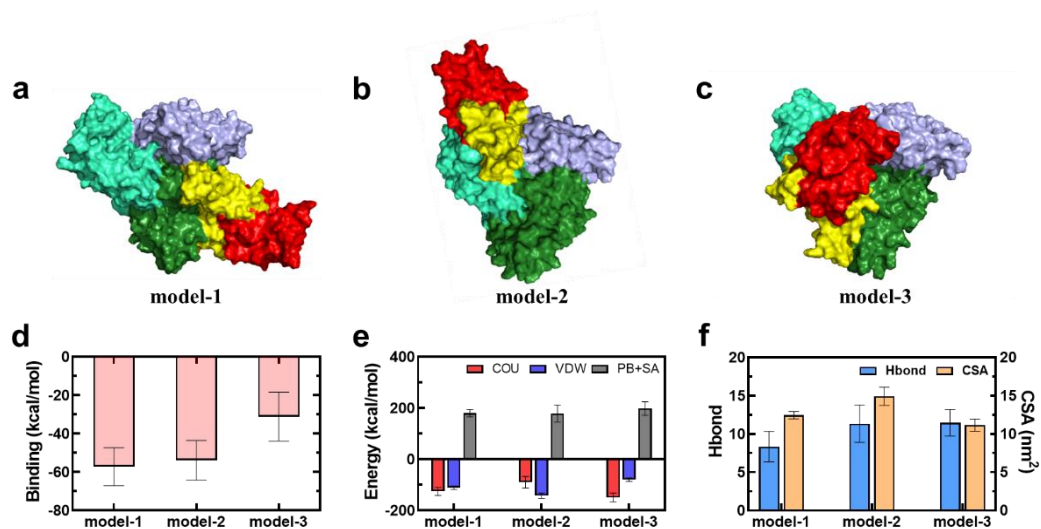

**Figure 4.** Interaction of HSA with SARS-CoV-2 RBD. Final snapshots illustrating the three typical modes in the HSA-RBD interaction (a-c) and corresponding binding energies (d), energy decomposition of the binding energy (e), and the hydrogen bonds and contact surface areas (f) between the HSA and RBD in the three modes.

We further investigated the effect of the HSA binding on the epitope in RBD. As shown in Figure 5a, the BSA of the key residues in the epitope was almost positive in model-1 and model-2, indicating that the epitope was largely occupied in these two models. Moreover, Figure 5b shows that nearly all ΔRMSF was negative in model-1, thus the flexibility of the RBD was also weakened; while half of the ΔRMSF was positive and the remaining was negative, thus the whole flexibility of the RBD was not affected in model-2. Considering that the binding energy was great in the two models, the HSA seems to have the potential of effectively neutralizing the RBD, particularly in model-1.

However, as we know, the antibody-dependent enhancement (ADE)[55, 56], in which the binding of a virus to the antibodies enhances its entry into host cells, is a potential risk to the vaccines or the antibody-based interventions and should be seriously considered. In this case, we also investigated the effect of the RBD binding on the active site of HSA (the FcRn receptor may recognize the site and metabolize the HSA[31, 57]). The BSA of some key resides in HSA was positive in model-1 and that of most residues was positive in model-2, indicating that the active sites may be occupied by the RBD. Moreover, the flexibility of the sites was greatly reduced in model-1 and model-2, thus the activity of these residues was weakened in both models. In general, there should not be the obvious 'ADE' of HSA, but its potential neutralization ability and safe use still need to be well evaluated in the experiment and particularly in the clinic trails.

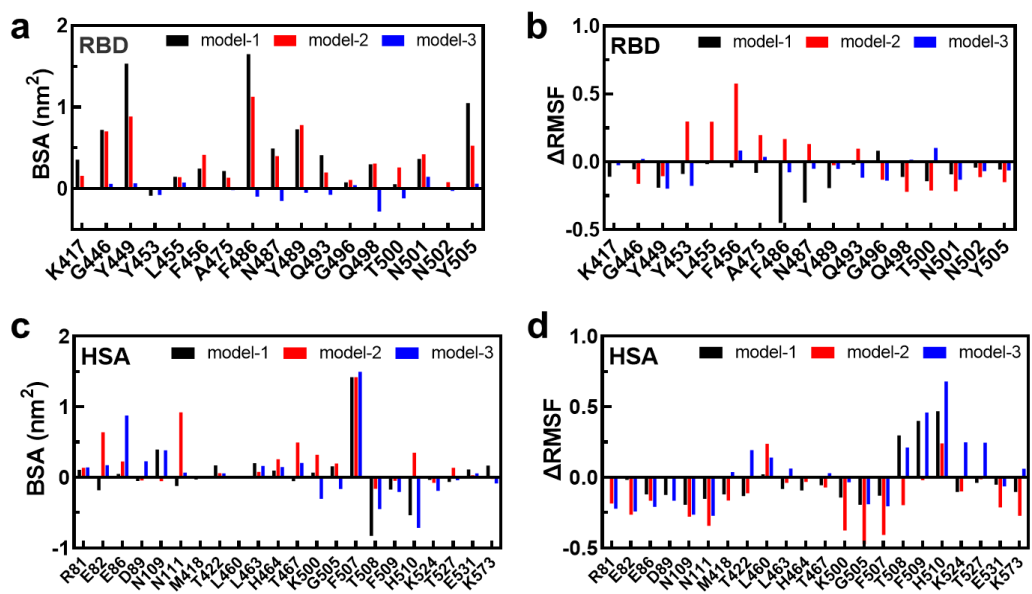

**Figure 5.** The BSA (a) and ΔRMSF (b) of the epitope site of RBD in the three models. The BSA (c) and ΔRMSF (d) of the active site of HSA in the three models.

### *3.3 Interaction of Apolipoprotein E (ApoE) with SARS-CoV-2 RBD*

Apolipoprotein is another important protein in the serum, which can bind to the lipids and interact with the lipoprotein receptors in the cell for the uptake and clearance of the lipids. In this case, we took the ApoE (that plays an important role in the transportation of cholesterol) as an example, and investigated its interaction with SARS-CoV-2 RBD (Figure 6 and Figure S3).

Similar to previous cases, three typical modes were also observed in the interactions (Figure 6a-c), where the RBD was mainly bound to the CT domain of ApoE with the RBM region (model-1), the junction between the CT domain and NT domain without (model-2) or with the RBM region (model-3). In particular, the lipid binding site in the CT domain of ApoE was occupied by the RBD in model-3 (Figure 6c). From the point view of the binding energy, the first two models were not stable since the binding energy was about -20 kcal/mol and -26 kcal/mol, and much weaker than the ACE2-RBD interaction. On the contrary, the binding energy in model-3 (~-34 kcal/mol) was much larger than that in the first two models, and comparable to that of ACE2-RBD interaction[50], which could be the result of more hydrogen bonds and hydrophobic contacts (Figure S3c-e). Notably, although the electrostatic energy and VDW energy were strong, the larger PB energy (probably due to a great number of polar residues at the binding site) led to the weak total binding energy in model-2.

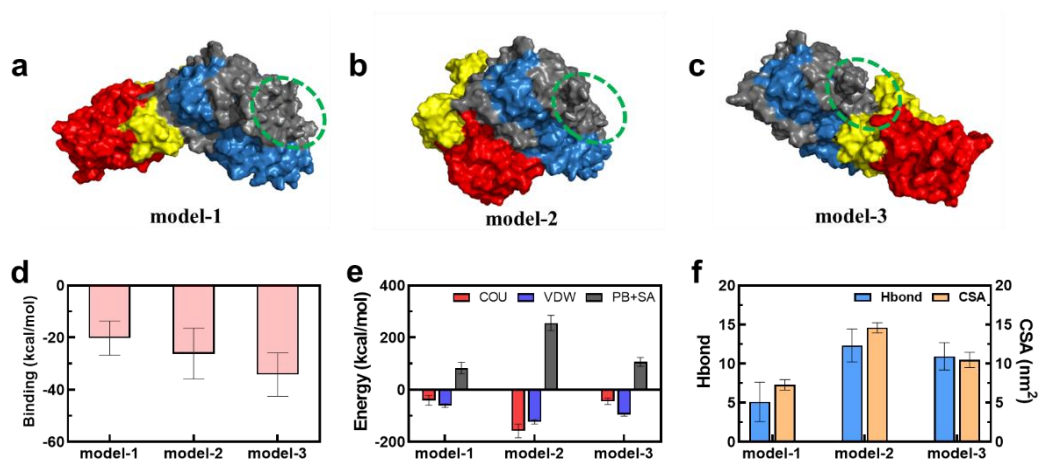

**Figure 6.** Interaction of ApoE with SARS-CoV-2 RBD. Initial snapshots illustrating the three typical modes in the ApoE-RBD interaction (a-c) and corresponding binding energies (d), energy decomposition of the binding energy (e), and the hydrogen bonds and contact surface areas (f) between the ApoE and RBD in the three modes. The lipid binding sites in CT domain are highlighted with the green dashed line.

We then investigated the effect of the ApoE binding on the epitope in RBD. As the ApoE was bound to the RBD with the RBM region in model-1 and model-3, the epitope would be greatly affected. As shown in Figure S4, the BSA for some key residues was highly positive in model-1 and model-3, while that was nearly zero in model-2. More importantly, most of ΔRMSF in model-3 was below zero, indicating that the flexibility of the epitope in RBD was great reduced. Notably, most of ΔRMSF in model-1 was positive, namely the activity of the epitope was enhanced due to the adsorption of ApoE. Since the binding energy was weak in model-1, such enhancement may further promote the substitution of ACE2 for ApoE (to SARS-CoV-2), which was very harmful. On the contrary, the ApoE was stably bound to the RBD, and blocked the epitope of the RBD effectively in model-3. In this sense, the model-3 may be a possible 'neutralization' mode of ApoE, but was the model-3 indeed useful for inhibiting the infection of SARS-CoV-2?

To answer this question, we further investigated the effect of the RBD binding on the active site of ApoE. As stated above, the ApoE was composed of two domains (i.e., NT domain and CT domain), where the active sites (residues 136-150)[58, 59] that can interact with the low-density-lipoprotein receptor (LDLR) on the cell (e.g., hepatocytes) were located on the NT domain, and the lipid binding sites (residues 260-299)[60] were located on the CT domain. In the absence of lipids, the active sites were buried by the other residues in ApoE. When the lipids were bound to the CT domain, the secondary structure of the NT domain was affected, which induced the exposure of the active sites. As a result, the cell could recognize the ApoE via the LDLR and internalize the adsorbed lipids along with the ApoE.

Similarly, Figure 7a shows that the BSA of the key residues in ApoE was great reduced in model-3, indicating that the active site became more exposed after its binding to RBD. Moreover, Figure 7b also shows that ΔRMSF was positive in model-3, indicating that the activity of the key resides was increased. As illustrated in Figure 7c-d, the binding to RBD indeed caused the change of secondary structure of ApoE, leading to a 15% loss of the helix in the CT domain of ApoE (Figure 7e-f). Thus, the SASA of the active site was great increased (compared to that in the free state) and became partially open. Generally, the above results collectively demonstrated that SARS-CoV-2 can induce the exposure and in turn activate the binding site of ApoE (to LDLR). In general, although the binding of ApoE could block the RBD-ACE2 interaction, the SARS-CoV-2 may enter the cell via activating the metabolic pathway of ApoE (instead of ACE2 pathway). Interestingly, Zhong et al.[61] recently found that the S1 subunit of SARS-2-S can bind to cholesterol and possibly to high-density lipoprotein (HDL) components to enhance viral uptake *in vitro*, and the HDL scavenger receptor B type 1 (SR-B1) facilitated ACE2-dependent entry of SARS-CoV-2. Moreover, we also noted a latest clinic data[62], where it was found that the level of low-density lipoprotein (LDL, including ApoE and ApoB) may be an indicator of the post treatment and the obvious decrease of LDL level was observed during the infection. Thus, our simulation results seemed to give a clear support on the above phenomena at the molecular level, which may shed light on the role of lipoproteins in the SARS-CoV-

2 infection.

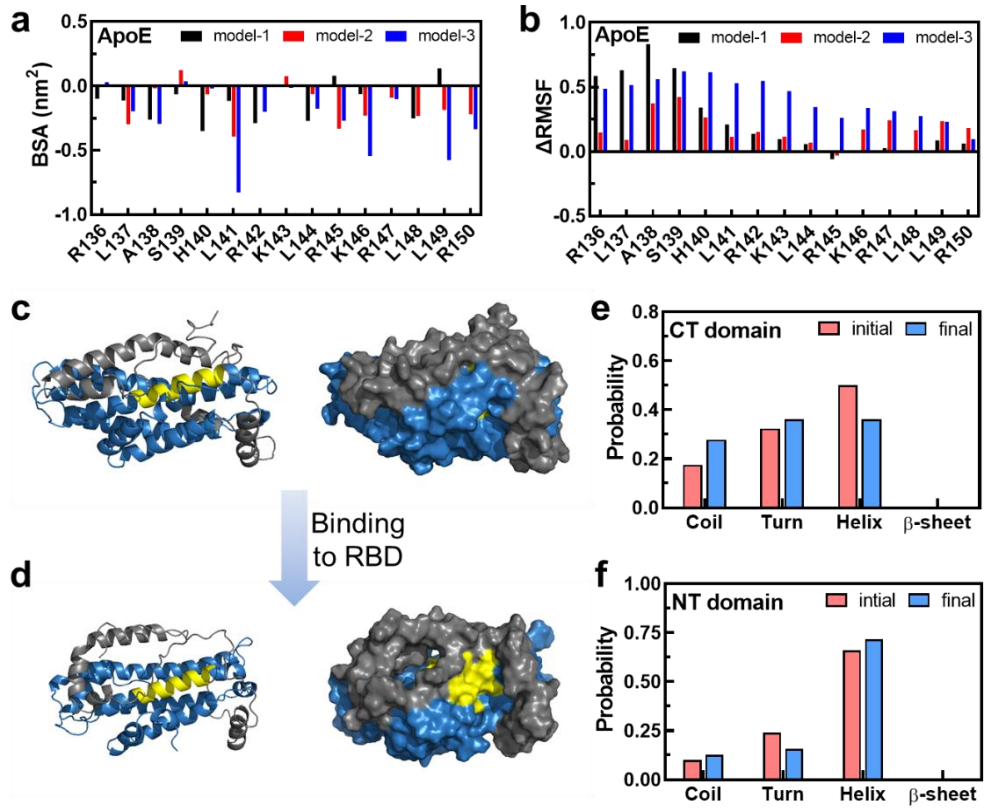

**Figure 7.** The BSA (a) and ΔRMSF (b) of the epitope site of ApoE in the three models. The cartoon (left) and surface (right) representation (c, d) showing the change of the second structure and the active site () of ApoE after the binding of SARS-CoV-2 RBD. The probability of the different secondary structures in the CT domain (e) and NT domain (f).

## 4. Conclusion

In summary, we have investigated the interactions of the SARS-CoV-2 RBD with three typical serum proteins. It is found that the non-specific IgM cannot effectively block the SARS-CoV-2-ACE2 interaction since its binding affinity with RBD is very weak. On the contrary, although HSA is not a specific 'antibody', it can bind to the SARS-CoV-2 RBD firmly due to its large space in the region junctions. More importantly, we found that the binding of SARS-CoV-2 RBD onto the ApoE can cause the exposure of the active site in ApoE; thereby the SARS-CoV-2 may enter the cell via the metabolic pathway of ApoE. In general, the present study clarifies the role of protein corona in the bio-behaviors of SARS-CoV-2 at the molecular level, and may shed some light on the development of the therapeutics against the COVID-19 disease in real applications.

## Conflict of interest

The authors declare no competing financial interest.

## Acknowledgments

This work is supported by the National Natural Science Foundation of China (Nos. 11874045 and 11774147).

## Author contributions

H.-M.D. and Y.-Q.M. conceived the idea and designed the project. Y-W.Y. performed the molecular-docking and the all-atom molecular dynamics simulation with the assistance of Y.-J.S. and M.W. H.-M.D. and Y-W.Y. analyzed the data and co-wrote the paper. All authors have proof-read the paper and approved its publication in the present form.

## Appendix A. Supplementary data

Supplementary Information accompanies this paper at http://doi.org/xxx.